\documentclass[
 preprint,onecolumn,
superscriptaddress,
showpacs,
showkeys,
nofootinbib,
 amsmath,
 amssymb,
 aps
]{revtex4-1}
\usepackage{bm}
\usepackage{graphicx}
\usepackage{epsfig}
\usepackage{subfigure}
\usepackage{booktabs}
\usepackage{amssymb}
\usepackage{amsmath}
\usepackage{lineno}
\usepackage{xcolor}
\usepackage{lipsum}
\usepackage[bookmarksopen=false]{hyperref}

\begin{document}
\title{The photon and a preferred frame scenario}
\author{Jakub Rembieli\'nski}
\affiliation{Chair of Theoretical Physics, University
of \L\'od\'z, ul.\ Pomorska 149/153, PL-90236 \L\'od\'z,
Poland.}
\email{jaremb@uni.lodz.pl}
\author{Jacek Ciborowski}
\affiliation{Faculty of Physics, University of Warsaw, Pasteura 5, PL-02093 Warsaw, Poland.}

\begin{abstract}
Structure of the space of photonic states is discussed in the context of a working hypothesis of existence of a preferred frame for photons.
Two polarisation experiments are proposed to test the preferred frame scenario.
\end{abstract}
\maketitle
\section{Introduction}
One of the most interesting questions in the present-day physics concerns
fundamental theory of space-time in the context of quantum gravity and a need of
an extension of the Standard Model. A possible implication of
contemporary approaches to  quantisation of gravity is breaking of the
Lorentz symmetry in the boost sector \cite{1,2}. In consequence, this leads to
existence of a preferred frame (PF) of reference.
A possibility that Nature might exhibit a preferred foliation of
space-time at its most fundamental level has attracted a serious attention since the last two decades.
One can mention Lorentz-violating extensions of the Standard Model
\cite{3,4,5,6} as well as new models of classical and quantum gravity e.g.
Einstein- aether \cite{7} and Ho\v{r}ava -- Lifshitz theories of gravity \cite{8}
(including  vacuum solutions in this model \cite{9}). Also worth mentioning
are the so called doubly-special relativity (DSR) theories \cite{10} which
are characterised by  modified dispersion relations for Lorentz violating
models. Almost all of the above theories predict new effects, however
suppressed by a power of the Planck scale. In particular, low energy
signatures of  Lorentz symmetry breaking in the photon sector include
vacuum birefringence. This is a consequence of asymmetry of the modified,
helicity dependent, dispersion relations for the photon. As a result,
rotation of the polarization plane is predicted, depending on the distance between the source
and the detector. Moreover, this effect also depends on a specific mechanism
of  Lorentz symmetry breaking by higher order differential operators \cite{11}.

In this paper, motivated by the preferred frame scenario, we consider
the problem of quantum description of the photon and its polarisation under
a minimal number of assumptions and from completely
different perspective than in the above mentioned dynamical theories. It is
shown that the presence of a PF of reference could results in some
polarisation phenomena caused by a specific structure of the Hilbert space
of photonic states.

\section{The relativistic approach to photonic states}

In the standard description of  photonic states one uses Hilbert space
${\cal H}$, which is a carrier space of a unitary, irreducible representation of the
inhomogeneous Lorentz group. The action of the Lorentz group in
${\cal H}$ is obtained by the Wigner-Mackey induction procedure \cite{12,13}. It can be
realised on the eigenvectors of the four-momentum operator and next extended
by linearity to the entire space. As a result one obtains:
\begin{equation}
U(\Lambda)|k,\lambda\rangle=e^{{\rm i}\phi(\Lambda,k)}|\Lambda k,\lambda\rangle,
\end{equation}
where $k=[k^\mu]$ is the photon four-momentum satisfying the dispersion relation
$k^2=k_\mu k^\mu=0$, $\Lambda$ is an arbitrary element of the homogenous Lorentz
group and the photon helicity $\lambda=\pm1$. Hereafter we will use the natural units
with $c=1$, $\hbar=1$. The inhomogeneous part of the Lorentz group is represented by
$e^{{\rm i}k^\mu \hat P_\mu}$, where $\hat P_\mu$ is the self-adjoint four-momentum
operator. The phase factor $e^{{\rm i}\lambda\phi(\Lambda,k)}$, representing the Wigner
little group element ${\cal L}_{\Lambda k}^{-1}\Lambda{\cal L}_k$ belonging to the
Euclidean group $E(2)$, realises its homomorphic unitary irreducible representation
which is isomorphic to the $SO(2)$ subgroup of $E(2)$
(an explicit form of the phase $\phi(\Lambda,k)$ can be found elsewhere \cite{14}). Here
${\cal L}_kq=k$, $q^T=\kappa(1;0,0,1)$ and $\kappa>0$.  The Lorentz-invariantly normalised
states $|k,\lambda\rangle$, $\langle
k,\lambda|p,\sigma\rangle=2k^0\delta^3({\bm k}-{\bm p})
\delta_{\lambda,\sigma}$, where $k^0=|{\bm k}|$, are identified with monochromatic, circularly
polarised. Therefore, the corresponding linearly polarised photonic states
have the form:
\begin{equation}
|\theta,k\rangle=\frac{1}{\sqrt{2}}(e^{{\rm i}\theta}|k,1\rangle+
e^{-{\rm i}\theta}|k,-1\rangle),
\end{equation}
where $\theta$ is the polarisation angle. Consequently, under Lorentz transformations (1)
the states (2) transform as:
\begin{equation}
U(\Lambda)|\theta,k\rangle=|\theta+\phi(\Lambda,k),\Lambda k\rangle,
\end{equation}
which means that linearly polarised states are transformed into linearly
polarized states related to a new phase.

\section{The preferred frame approach to photonic states}

Our aim is to apply the Wigner-Mackey construction to the case when one
inertial frame is physically distinguished. We will assume
the Lorentz covariance under transformations between inertial frames.
Obviously, this assumption does not exclude the case when the Lorentz
symmetry is broken because we deal with {\em passive\/} space-time transformations.
From the point of view of an inertial observer, the preferred frame has a
time-like four-velocity $u^\mu$ i.e.\ ${u^0}^2-{\bm u}^2=1$. The observer's frame will
be denoted as $\Sigma_u$ while the PF corresponds to $u^T_{PF}=(1;0,0,0)$. It can be seen  that the
working hypothesis that Nature distinguishes a preferred inertial frame of reference is
nontrivial in the photonic sector only if the monochromatic photonic states are frame-dependent
i.e.  they depend not only on $k^\mu$ but also on $u^\mu$. Hereafter we will denote them as
$|k,u,\lambda\rangle$. The Hilbert space of the observer in $\Sigma_u$ will be denoted by
${\cal H}_u$.  The family of Hilbert spaces ${\cal H}_u$ form a fiber bundle corresponding to
the bundle of inertial frames $\Sigma_u$ with the quotient manifold $SO(1,3)/SO(3)\sim {\mathbb R}^3$
as the base space. As in the standard case, the Hilbert space ${\cal H}_u$
is spanned by eigenvectors of the four-momentum operator but these base
vectors are $u^\mu$-dependent. To apply the Wigner-Mackey construction to this case we should
relate each pair of four-vectors $(k,u)$ with the "standard" pair $(q,u_{PF})$
and determine the stabiliser of the standard pair. It is obvious that the
little group of the pair $(q,u_{PF})$ is $O(2)\sim E(2)\cap O(3)$.  Moreover, the pair
$(k,u)$ can be obtained from the standard pair $(q,u_{PF})$ by the sequence of
Lorentz transformations $L_uR_{\bm n}$, where $L_u$ is the Lorentz boost transforming
$u_{PF}$ into $u$ and $R_{\bm n}$ is the rotation of $q$ into four-vector
$\kappa(1;{\bm n})$, provided the unit vector ${\bm n}$ is equal to:
\begin{equation}
{\bm n}={\bm n}(k,u)=\frac{1}{uk}\left({\bm k}-\frac{|{\bm
k}|+uk}{1+u^0}\right),
\end{equation}
where $uk=u^\mu k_\mu=\kappa$. Applying the Wigner-Mackey procedure to the base vectors
$|k,u,\lambda\rangle$ in the manifold of Hilbert spaces ${\cal H}_u$
we obtain the unitary action of the Lorentz group of the form:
\begin{equation}
U(\Lambda)|k,u,\lambda\rangle=e^{{\rm i}\varphi(\Lambda,k,u)}|\Lambda k,\Lambda u,\lambda\rangle,
\end{equation}
where $e^{{\rm i}\varphi(\Lambda,k,u)}$ is the phase representing the
Wigner rotation:
\begin{equation}
W(k,u,\Lambda)=(L_{\Lambda u}R_{{\bm n}(\Lambda k,\Lambda u)})^{-1}\Lambda
L_u R_{{\bm n}(k,u)},
\end{equation}
belonging to the subgroup $SO(2)$. The Wigner rotations satisfy the group
composition law of the form:
\begin{equation}
W(k,u,\Lambda_2\Lambda_1)=W(\Lambda_1k,\Lambda_1u,\Lambda_2)W(k,u,\Lambda_1).
\end{equation}
Now, by means of (5) the linearly polarised states:
\begin{equation}
|\theta,k,u\rangle=\frac{1}{\sqrt{2}}(e^{{\rm i}\theta}|k,u,1\rangle+
e^{-{\rm i}\theta}|k,u,-1\rangle)
\end{equation}
transform under the Lorentz group action unitarily from ${\cal H}_u$
into ${\cal H}_{\Lambda u}$ according to the transformation law:
\begin{equation}
U(\Lambda)|\theta,k,u\rangle=|\theta+\phi(\Lambda,k,u),\Lambda k,\Lambda u\rangle .
\end{equation}
Furthermore, the ideal polariser, regarded as a quantum observable in ${\cal H}_u$,
can be defined as the projector:
\begin{equation}
\Pi^{u,\varTheta}_{\Omega({\bm n})}=\int_{{\mathbb R}_+\times\Omega}\frac{d^3{\bm p}}{2|{\bm p}|}
|\varTheta,p,u\rangle\langle\varTheta,p,u|=\textstyle{\frac{1}{2}}\int_0^\infty|{\bm p}|d|{\bm p}|
\int_{\Omega({\bm n})}d\Omega|\varTheta,p,u\rangle\langle\varTheta,p,u|,
\end{equation}
where $\frac{d^3{\bm p}}{2|{\bm p}|}$ is the Lorentz invariant measure, the polarisation angle
$\theta$ is fixed, while $\Omega({\bm n})$ is a solid angle around a fixed direction ${\bm n}$.
A photon, in order to be detected, should have his momentum direction in the solid angle
$\Omega({\bm n})$; otherwise it cannot pass through the polariser. Indeed,
applying $\Pi^{u,\varTheta}_{\Omega({\bm n})}$ to a linearly polarised state $|\theta,k,u\rangle$
we find that:
\begin{align}
\Pi^{u,\varTheta}_{\Omega({\bm n})}|\theta,k,u\rangle=
\begin{cases}
\cos(\varTheta-\theta)|\varTheta,k,u\rangle &\hbox{\rm if }{\bm k}\in\Omega({\bm n}),\cr
0 &\hbox{\rm if }{\bm k}\notin\Omega({\bm n}).
\end{cases}
\end{align}
It is evident that this observable satisfies the quantum Malus law \cite{15}.
Indeed, the probability $p(\theta,\varTheta)$ of finding a linearly
polarised photon in the polarised state determined by the polarisation angle
$\varTheta$ has the form:
\begin{equation}
p(\theta,\varTheta)=\cos^2(\varTheta-\theta).
\end{equation}
It follows from the definitions of the phase factors in eqs. (1) and (5) that
the change of the polarisation angle is different  in the presence or absence of a PF.
The difference is expressed by a nontrivial phase shift
$\Delta\phi=\phi(\Lambda,k,u)-\phi(\Lambda,k)$.  In principle, this
"geometric" phase shift can be explicitly calculated
by means of (6) as well as measured.

\section{Two direct experiments}

In order to illustrate the above result, let us analyse two possible experiments.
Firstly, let us imagine two observers in inertial frames $\Sigma_u$ and $\Sigma_{u'}$
related by the Lorentz transformation $\Lambda(V)$ determined by the velocity ${\bm V}$
along the photon momentum direction i.e.\ ${\bm k}\|{\bm V}\|{\bm k}'$
as shown in Fig.\ 1.
\begin{figure*}
\centering
\includegraphics[width=12cm]{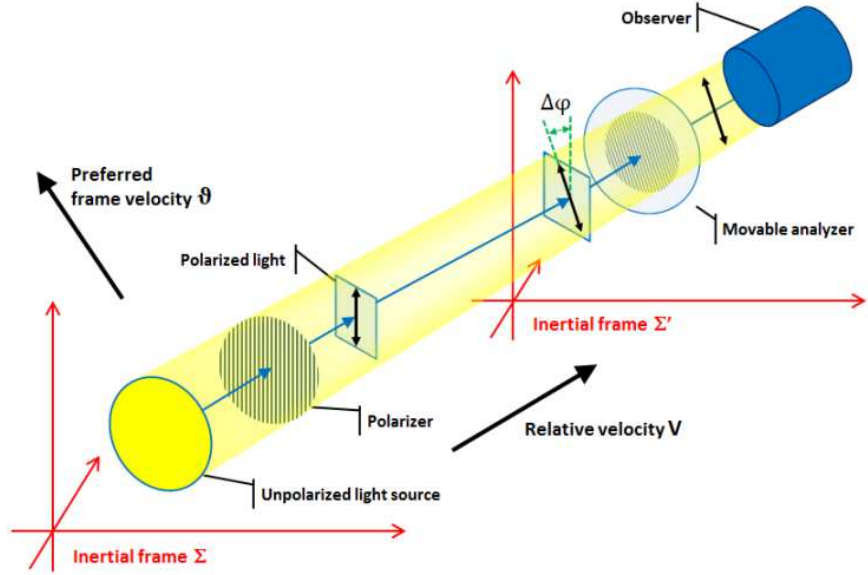}
\caption{A schematic presentation of the polarisation experiment for inertial
observers in relative motion.}
\end{figure*}
Linearly polarised photons are send by one observer and detected by
the other. In that case the standard phase shift is equal to zero \cite{14}.
On the other hand, the phase shift $\phi({\bm V},k,u)$ in this configuration,
calculated from (6), is given by:
\begin{equation}
\phi(V,\vartheta,\chi)=\arcsin\frac{V\vartheta\sin\chi}
{\sqrt{2(1+\sqrt{1-V^2})(1+\sqrt{1-\vartheta^2})(V\vartheta\cos\chi+\sqrt{1-V^2}
\sqrt{1-\vartheta^2}+1)}}.
\end{equation}
Here the  velocities $V=\pm|{\bm V}|$ and $\vartheta=|{\bm\vartheta}|$
are expressed  in the units of $c$ and $\chi$ is the angle between ${\bm k}$ and
the preferred frame velocity ${\bm\vartheta}=\frac{{\bm u}}{u^0}$,
as seen by the observer in the frame $\Sigma_u$.  In this case the phase difference
in general does not vanish, $\Delta\phi=\phi(V,k,u)\neq0$.  One can  see that for ${\bm\vartheta}$
parallel to the photon momentum ($\chi=0$), the phase shift is zero whereas for
${\bm\vartheta}$ perpendicular to the photon momentum ($\chi=\frac{\pi}{2}$)
it reaches the maximal value. The dependence of $\phi(V,k,u)$ as a function of the
relative velocity $V$ with the choice of the CMBR frame as the PF for
$\chi=\frac{\pi}{2}$ is presented in Fig.\ 2. In this case the observer in $\Sigma_{u'}$
measures the rotation of the polarisation plane of the photon in comparison to
the polarisation plane of the initial light beam.
\begin{figure*}
\centering
\includegraphics[width=12cm]{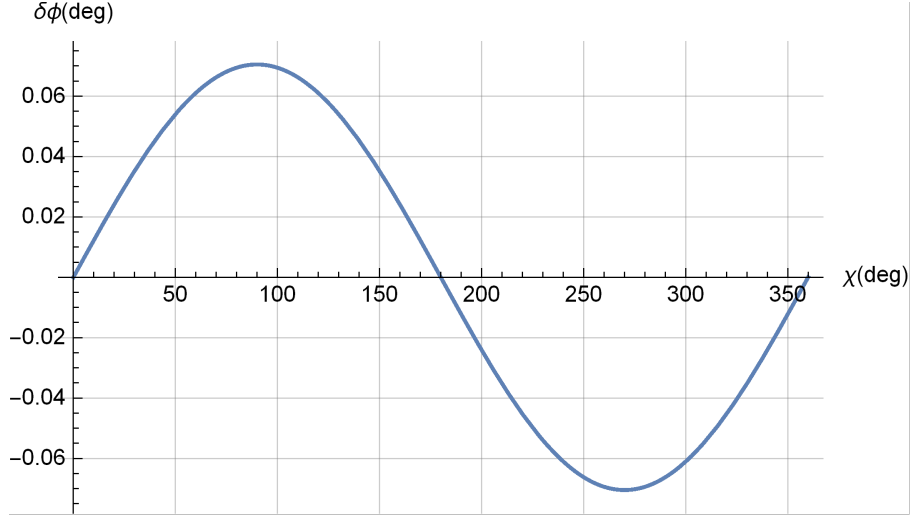}
\caption{The phase $\phi$ as a function of the boost velocity $V$ for PF
identified with the CBMR frame.  The photon momentum ${\bm k}$
is chosen as perpendicular to the preferred frame velocity
${\bm\vartheta}$ ($\chi=\frac{\pi}{2}$).}
\end{figure*}

Now, let us consider another possible consequence of the influence of the PF  on the photon polarisation.
If we apply the  formula (6) to the case of a rotation $R(\delta)$ around the photon momentum in a fixed frame we obtain:
\begin{equation}
\varphi[R(\delta),k,u]=2\arctan\frac{\sqrt{1-\vartheta^2}+
[(1-\sqrt{1-\vartheta^2})\cos\chi-
\vartheta]\cos\chi}{(1-\vartheta\cos\chi)\cot
\frac{\delta}{2}+[(1-\sqrt{1-\vartheta^2})\cos\chi-
\vartheta]\sin\chi}.
\end{equation}
Taking into account that the corresponding phase shift for the standard case
is exactly $\delta$ \cite{13}, the phase shift difference is of the form
$\Delta\phi=\delta-\varphi[R(\delta),k,u]$.  For $\vartheta\ll1$
we obtain:
\begin{equation}
\Delta\phi\simeq 2\vartheta\frac{\sin\chi\tan^2\frac{\delta}{2}}
{1+\tan^2\frac{\delta}{2}},
\end{equation}
which is depicted in Fig. 3, where the preferred frame is identified with the CMBR frame.
In this case  an anomalous correction to the classical Malus law is present.
\begin{figure*}
\centering
\includegraphics[width=12cm]{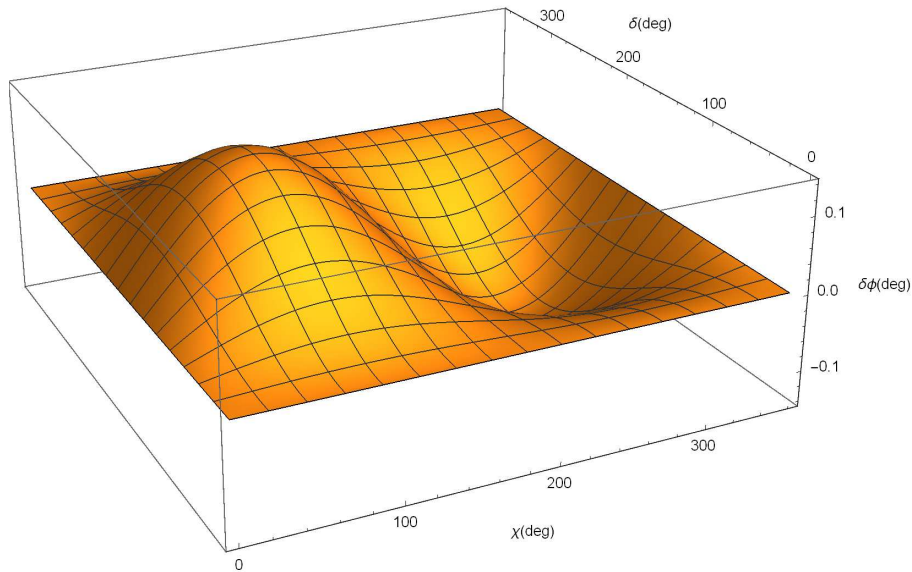}
\caption{The phase shift difference $\Delta\varphi$ as a function of the rotation
angle $\delta$ around of the photon velocity and of the angle between preferred
frame velocity and the photon velocity for PF identified with the CBMR frame.}
\end{figure*}

\section*{Conclusions}

We have analysed possible consequences of existence of a preferred frame for the photon
assuming the standard dispersion relation  and the Lorentz
covariance realised \textit{via} passive transformations. The crucial assumption is that
the quantum state of the photon, as seen by an inertial observer, depends on
the observed velocity of the PF. Under these assumptions, by
means of the Wigner- Mackey method of the induced representations, we have
constructed the space of single photon states. We also have defined an ideal
polariser, regarded as a quantum observable in the Hilbert space of states,
satisfying the quantum Malus law. For linearly polarised states we have
obtained a difference between the Wigner phases for the standard case (absence
of PF) and for the theory with a PF.
This difference manifests as an additional rotation of the polarisation plane of
linearly polarised photons. Such optical effect has rather geometrical than
dynamical nature and is different from the vacuum birefringence which
appears in models with modified energy-momentum dispersion relations of
the photon. The effect is  independent of the photon energy (frequency) and of the distance
between the source and observer but instead  depends on relative velocity of
the reference frames, velocity of PF and on relative configuration of these velocities.
Two direct experiments were proposed to test the PF  hypothesis.
In one of them the predicted effect, if exists, can be observed as a
deviation from the classical Malus law.
\section*{Acknowledgements}
The authors would like to thank Pawe{\l} Caban, Krzysztof
Kowalski, Kordian Smoli\'nski and Marek \.Zukowski for comments and discussions
as well as Pawe{\l} Horodecki for inviting Jakub Rembieli\'nski to give lectures
related to this work on the conference {\em Quantum Phenomena: Between the Whole
and the Part}, and {\em Quantum Foundations and Beyond}, held in Sopot, Poland in September
2016, and December 2017, respectively and supported by the
John Templeton Foundation. This work has been supported by the Polish National
Science Centre under the contract 2014/15/B/ST2/00117 and the University of Lodz grant.

\end{document}